# Sigmoid function based intensity transformation for parameter initialization in MRI-PET Registration Tool for Preclinical Studies

Hiliwi Leake Kidane, Stephanie BRICQ, Alain LALANDE
Laboratory Le2i, Universite Bourgogne - Franche-Comte,
21000 Dijon, France,

***Abstract***. *Images from Positron Emission Tomography (PET) deliver functional data such as perfusion and metabolism. On the other hand, images from Magnetic Resonance Imaging (MRI) provide information describing anatomical structures. Fusing the complementary information from the two modalities is helpful in oncology. In this project, we implemented a complete tool allowing semi-automatic MRI-PET registration for small animal imaging in the preclinical studies. A two stage hierarchical registration approach is proposed. First, a global affine registration is applied. For robust and fast registration, principal component analysis (PCA) is used to compute the initial parameters for the global affine registration. Since, only the low intensities in the PET volume reveal the anatomic information on the MRI scan, we proposed a non-uniform intensity transformation to the PET volume to enhance the contrast of the low intensity. This helps to improve the computation of the centroid and principal axis by increasing the contribution of the low intensities. Then, the globally registered image is given as input to the second stage which is a local deformable registration (B-spline registration). Mutual information is used as metric function for the optimization. A multi-resolution approach is used in both stages. The registration algorithm is supported by graphical user interface (GUI) and visualization methods so that the user can interact easily with the process. The performance of the registration algorithm is validated by two medical experts on seven different datasets on abdominal and brain areas including noisy and difficult image volumes..*

## 1. INTRODUCTION

The use of small animal models in preclinical studies constitutes an integral part of testing new pharmaceutical agents and exploring new biological functions [1]. The mouse and the rat are the most widely used animals in medical research. Each medical imaging modality has its own advantages and limitations and acquired information is actually complementary between them [1]. Consequently, multimodal approach is used to reveal both anatomical (MRI and CT) and functional (PET, SPECT, or optical imaging) information. Alignment of these images requires the use of multimodal registration methods [3]. Among all combinations of modalities, PET-CT and PET MRI are the most mature combinations. However PET-CT has shortcoming due to the significant radiation dose to the small animal contribute by CT and MRI offers better contrast among soft tissues compared to CT [4]. As a result, PET-MRI which offers the combination of high resolution, soft tissue, anatomical information of MRI, and high sensitivity of PET[1] is a promising combination in preclinical research and will certainly progress to clinical application[4]. The common small animal biological studies involving PET and MRI acquisitions are tumour imaging, brain imaging and cardiovascular imaging [5].

Some works has been proposed to register MRI-PET images of small animals. Vaquero et al. [11] investigated the MRIPET registration algorithms developed by Woods at al.[12] and Collignon et al.[7] to register PET images to CT or MR images of the rat skull and brain. The latter was found to be more robust algorithm than the former method. Hayakawa et al. [13] modified the algorithm proposed by [9] to register PET and MR images of rat brains. Bernier et al. [14] proposed parallel multi-resolution and PCA initialization for MRI-PET registration of small animal bones. As many of the previous works on small animal MRI-PET registration are focused on the head and bones, i.e a rigid body registration, the non-rigid registration problem remains more open and an active area of research.

In this paper, a complete tool MRI-PET registration based on a two-level hierarchical registration steps supported by GUI interface and visualization is developed. In PET images, only the low intensities reveal the anatomical structure in MRI scan. Consequently, focusing on the PET range of intensity which reveals the anatomic structure in MRI will help to align the images perfectly. We apply a non-linear intensity transformation to the PET volume to enhance the contrast of the low intensity. The computation of initial parameters using PCA and the process of finding the optimal global affine transformation is performed using the intensity transformed PET. Then, the original

PET volume is transformed using the final optimal global affine transformation matrix and given to the local registration as input. Moreover, we develop a visualization and GUI support for the registration algorithm so that the user can interact with the registration to select volume of interest and visualize the input/outputs files.

## 2. METHOD

The flow-diagram of the proposed intensity transformation based hierarchical semi-automatic registration algorithm is shown in figure 1. The details of the block in the flow-diagram are discussed in the following sections.

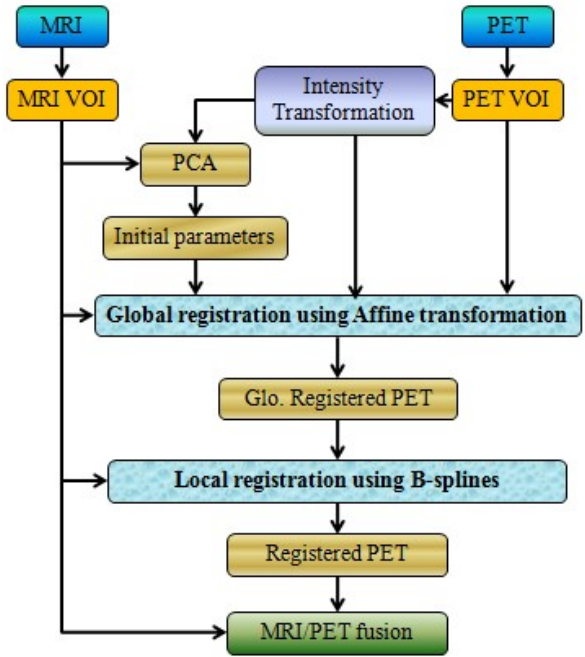

Figure 1 Flow diagram of the proposed algorithm

### 2.1. Volume of interest (VOI) selection

Since the small animals are not cooperative like humans, anaesthesia is used throughout the acquisition session to keep the animal in the same position, i.e the imaging session is governed by anaesthesia and it is limited. This limited time is enough to take PET scan of large part of body. In contrast, as different sequences and weighting are considered during the acquisition of MRI, it is difficult to take scan of large part of the body for each types of sequence and weighting. Consequently, the PET volume is always larger than MRI.

As the whole body images of small animal contain many articulated joints and the PET volume lacks spatial details, it is difficult to initialize the registration without avoiding the non-overlapping region. Moreover, the PCA is not useful if the two volumes do not refer to the same body. Selecting VOI will help to avoid unwanted objects present in the image volumes from affecting the registration outcome.

For better output and not to miss slices during the selection of VOI, we introduce a method to compare the slice thicknesses in both volume and add appropriate slice to the volume with thinner slice as shown in Fig. 2.

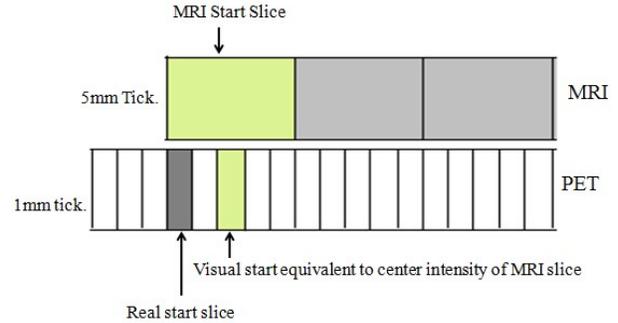

Figure 2 Flow diagram of the proposed algorithm

### 2.2. Intensity transformation of PET

While registering PET image with MRI, it is important to focus on the range of intensities which reveal the anatomic structure in the MRI scan. Normally, only the low intensities in the PET volume represent the full anatomical structure in MRI scan. Increasing the dynamic range of the low intensity will support to the PCA to compute the right centroid and principal axis by maximizing the density of 1s' in the PET volume to comparable level with the density of 1s' in the MRI volume. This can be alleviated by employing a sigmoid function [17]. It is a non-linear mapping which maps a specific range of intensity values into a new intensity range by making a very smooth and continuous transition in the borders of the range. Sigmoid function is given by:

$$I' = (Max - Min) \frac{1}{1 + e^{-\left(\frac{I-\beta}{\alpha}\right)}} - Min$$

Where I is the input intensity and I' is the transformed intensity, Max and Min the maximum and minimum of the expected output image, α defines the width of the input intensity range, and β defines the intensity around which the range is cantered [16].

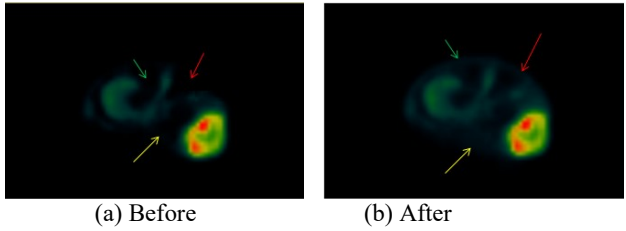

(a) Before          (b) After

Figure 3 PET slices before and after intensity transformation

### 2.3. Principal component analysis (PCA)

PCA is a technique that computes a linear transformation to map a high dimensional space into a lower dimensional space. The basis of PCA is computed by the Eigen decomposition of the data covariance matrix [18]. The idea of PCA initialization is derived from the theory of rigid body where a rigid body is uniquely located by knowledge of its center of mass (centroid) and its orientation (rotation) with respect to its center of mass[18]. PCA produces a single best line in such a way that the sum of the squares of the perpendicular distances from the sample points to the line is a minimum. The first principal component is the variable defined by the line of best fit which indicates the greatest amount of variation whereas the second principal component is the variable defined by the line that is orthogonal with the first and the center of the data set is the intersection of the two axes [19]. We have implemented PCA to find the centroid and orientation of image PET and MRI volumes to initialize the translation and rotation as described by Lu and Chen[19].

### 2.4. Global Registration using Affine transformation

The global transformation model describes the overall motion of the animal body. An affine transformation parameterized by 12 degrees of freedom (DOF) is proposed for the global motion. Since the two modalities have different resolution, the rigid registration with 6 DOF is not sufficient to overcome the global motion. For 3-D images, the affine transformation can be written as:

$$T_G(x,y,z) = \begin{pmatrix} \theta_{11} & \theta_{12} & \theta_{13} \\ \theta_{21} & \theta_{22} & \theta_{23} \\ \theta_{31} & \theta_{32} & \theta_{33} \end{pmatrix} \begin{pmatrix} x \\ y \\ z \end{pmatrix} + \begin{pmatrix} \theta_{14} \\ \theta_{24} \\ \theta_{34} \end{pmatrix}$$

Where the coefficients, $\theta()$, parameterize the 12 DOF of the transformation for 3-DOF rotations (R), 3-DOF transformations, 3-DOF scaling (S) and 3-DOF shearing (H). The initial rotation and translation are computed using PCA and the initial scaling and shearing are assumed to be identity, i.e initially there is no scaling difference between the volumes and there is no shearing problem.

### 2.5. Local registration using B-splines

The scope of the rlobal registration using Affine transformation is to align the two volumes globally. However, there is local deformation due to breathing and uncontrolled movement in the lower abdomen of the small animals during acquisition. A non-rigid Cubic B-spline free-form deformation (FFD) is used for the local registration. The motivation to choose Cubic B-splines for the local deformable registration is that, B-splines is the most adequate basis function to represent the deformation with very small overlap which makes it faster and reduce the interdependency between the parameters as demonstrated by Kybic and Unser [20].

### 2.6. Interpolation

Interpolation is a method of constructing new data points within the range of a discrete set of known data points. Image volumes are sampled at discrete grid points, P and when the image's grid points are transformed to align with other image, the grid point does not coincide with the other grid points. Hence interpolation must be applied to calculate the intensity values at the new grid points using the information from neighboring pixel or voxel grid positions. Different interpolation methods are proposed for image registration. Among them, we applied B-spline interpolation because it is the most effective interpolation scheme having the superior performance than any other polynomial basis function of the same order and is highly recommended for multi-resolution registration strategy [23] [24].

### 2.7. Similarity metric

Normalized mutual information (NMI) [25] is used as a similarity metric in both stages of registration. Mutual information is a measure of the amount of information one random variable contains about another. In the context of image registration, image intensity is a random variable and MI measures how much image intensity in one image tells about image intensity in the other image and is defined in terms of entropy [26].

Entropy is self information of a random variable or a measure of uncertainty of random variable. The mutual information of image X and Y is given by:

$$MI(X,Y) = H(X) + H(Y) - H(X,Y)$$

Where *H(X)*, *H(Y)* demote the marginal entropies of *X*, *Y* and *H(X,Y)* denotes their joint entropies. Let $P_X(x)$ and $P_Y(y)$ are probability distribution of intensity values of x and y of image X and Y, then the marginal entropies are given as:

$$H(X) = -\sum_x P_x(x)\log(P_x(x))$$

and

$$H(Y) = -\sum_y P_Y(y)\log(P_Y(y))$$

Similarly the joint entropy *H(X,Y)* of a pair of random variables *(X,Y)* with a joint probability density function $P_{XY}(x,y)$ can also be defined as:

$$H(X,Y) = -\sum_{x,y} P_{XY}(x,y)\log(P_{XY}(x,y))$$

If both images are aligned, the mutual information is maximized. It has been shown by Studholme et al.[25] that mutual information is not independent of overlap between two images. To overcome any dependency on the amount of overlap, the authors suggested the use of normalized mutual information (NMI) as measure of image alignment. The NMI is given as:

$$NMI = \frac{H(X) + H(Y)}{H(X,Y)}$$

In practice, direct access to the marginal and joint probability densities is not possible and hence the densities must be estimated from the image data. The two most efficient techniques used for probability density estimation are discrete joint histogram and Parzen windowing [16] [23]. In case of discrete joint histogram, the marginal and joint probability densities are computing by counting the number of occurrence of each intensity value in the images. This method does not allow the similarity metric to be explicitly differentiated and can only be used with non-gradient based optimization methods. Whereas, in Parzen windowing, the marginal and joint densities are estimated by constructing a continuous density function by superimposing kernel functions centered on the intensity samples obtained from the images. Parzen windowing provides a continuous joint histogram which is a derivative function, so that gradient based optimization method can be applied in the registration process [27]. In this project, mattes mutual information which uses Parzen windowing for estimation of the density distributions implemented in ITK [16] is used.

### 2.8. Optimization

Optimization algorithms find the optimal transformation parameters that can align volumes by minimizing the negated mutual information. Since both global and local registrations are initialized well, a derivative optimization method is used. Then, a regular step gradient-descent is selected to optimize the mutual information of the affine global registration [29][10]. For the local registration where the B-spline transform has a high dimension of parameter space, a Limited memory Broyden-Fletcher Goldfarb-Shannon with bounds (LM-BFGS-B)[30] is used.

### 2.9. Registration, Visualization and GUI tools

To develop the tool, C++, ITK, VTK and Qt are used.

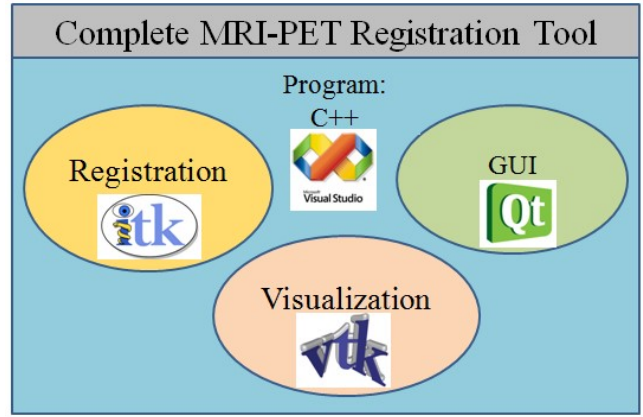

Figure 4 Tools used for development

The registration algorithm is developed using Insight Toolkit (ITK) [16]. Though, ITK provides advanced algorithms for performing image registration and segmentation, it does not provide support to perform image visualization, nor does it offer any graphical user interface (GUI) framework. Consequently, the Visualization Toolkit (VTK) [31] which is an open-source, freely available software system for 3D computer graphics, image processing and visualization is integrated with ITK for visualization purpose. The GUI is developed by integration Qt, another cross-platform application framework that is widely used for developing application software with a graphical user interface (GUI). The programs are written using Visual Studio C++. The developed GUI and visualization is given in Fig. 4.

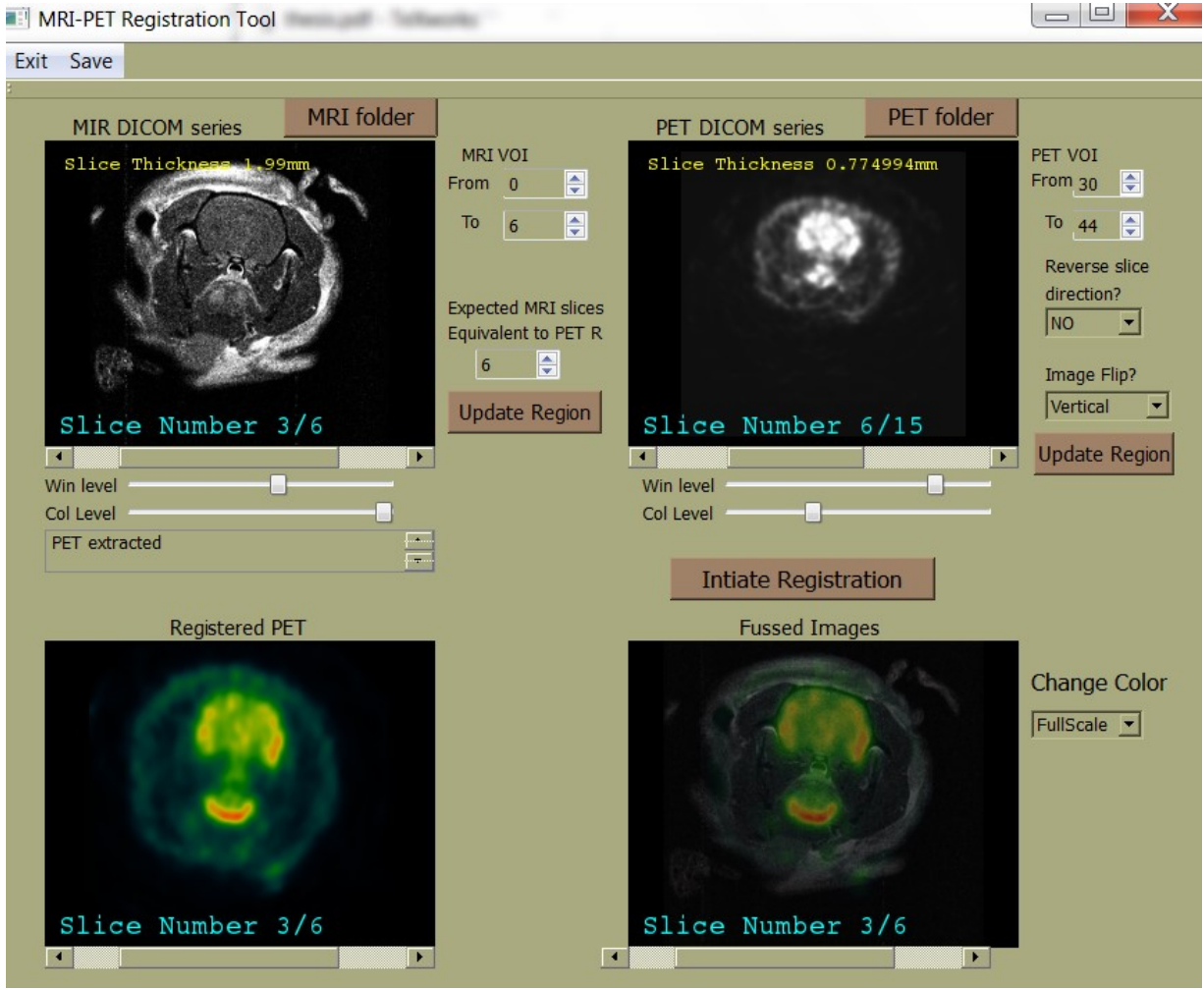

**Figure 5** GUI interface of the registration tool

## 3. EXPERIMENTS AND RESULTS

The developed semi-automatic MRI-PET registration algorithm is tested and validated using seven different datasets. The experiment was carried out using processor of Intel(R) Core(TM) i3-2350M CPU@ 2.30GHz 2.30GHz, RAM 4.00GB(2.70 usable) running in Windows 7 32-bit.

### 3.1. Experiment dataset

The dataset used to investigate the performance of this developed algorithm were obtained from preclinical imaging laboratory at Dijon in the framework of the IMAPPI (Integrated Magnetic resonance And Positron emission tomography in Preclinical Imaging) project and consists of brain and abdominal MRI and PET images of rat and mice. The brain scan dataset used was deformed both globally and locally. On the other hand, the abdominal scan datasets used contain slightly deformed and noise volumes. All the dataset were axial images and their detail size is given in Table I.

| Subject | Test | Modality | Dimension | Voxel size (mm) |
|---------|------|----------|-----------|-----------------|
| Abdomen | Test1 | MRI T1 | 256x256x7 | 0.27x0.27x3 |
| | | PET | 176x176x48 | 0.39x0.39x0.3875 |
| Brain | Test2 | MRI T1 | 256x256x6 | 0.2x0.2x2.01 |
| | | PET | 176x176x12 | 0.39x0.39x0.7749 |
| Abdomen | Test3 | MRI T2 | 256x256x7 | 0.12x0.12x3.0 |
| | | PET | 175x175x23 | 0.39x0.39x0.775 |
| Abdomen | Test4 | MRI T1 | 256x256x7 | 0.27x0.27x3 |
| | | PET | 176x176x25 | 0.39x0.39x0.775 |
| | Test5 | MRI T1 | 256x256x7 | 0.27x0.27x3 |
| | | PET | 176x176x25 | 0.39x0.39x0.775 |
| | Test6 | MRI T2 | 256x256x7 | 0.12x0.12x1.5 |
| | | PET | 175x175x25 | 0.39x0.39x0.775 |
| Brain | Test7 | MRI T1 | 256x256x16 | 0.2x0.2x1.99 |
| | | PET | 176x176x6 | 0.39x0.39x0.78 |

### *3.2. Experiment Results*

The performance of the algorithm is assessed using two brain and five abdominal datasets. The registration success evaluation in percentile is given in Figure 6. The maximum score is for abdomen test where it has a PET volume with the minimum slice thickness of all. As indicated in Table I, the PET slice thickness for test-1 is 0:3875 which is half of the slice thickness of other datasets.

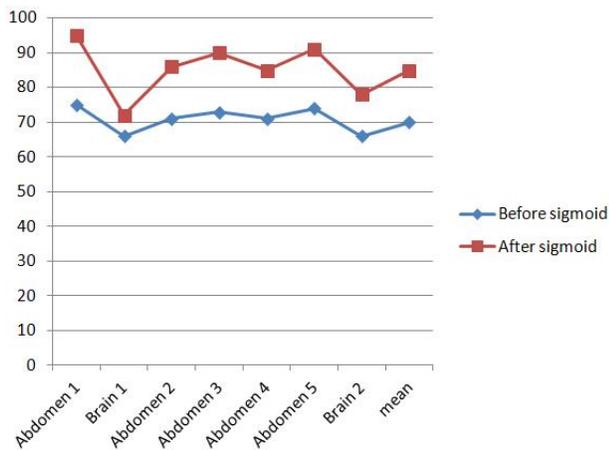

Figure 6 Registration successes of the 7 datasets in %

The minimum score is for the brain test which is the most difficult of all the other datasets. In this dataset, there is abrupt change or deformation in the MRI scan from one slice to the other which does not exist in the PET slices. However, the overall result of the registration is promising with success rate 85%. The optimization time is also reduced by 23% due to the introduction robust initialization using the intensity transformation. A sample visual registration result of brain and abdominal are given in Figures 7 and 8 respectively.

The first columns of the figures contain the original MRI slices. In the second columns of the figures, a corresponding original PET slices is provide. The slices in the third columns of Figures 7 and 8 are the PET slices after registration. Finally, the fusion of both the MRI and PET slices is provided in the fourth columns of the figures. Both visual results show that the slices are well aligned/fused. The lower score of the brain registration is due to the noisy ear visible in the MRI scan only.

## CONCLUSION

In this paper, we have presented implementation of a complete tool allowing semi-automatic MRI-PET registration for pre-clinical studies. The sigmoid function based non-uniform intensity transformation to the PET volume boosts the contribution of the low intensities to compute the initial parameters improve the overall registration. The performance of the registration tool is very promising. The mean registration success of the 7 datasets is above 85%. The mean processing time of the above datasets is 12sec. Moreover, a comparison of the registration before and after using intensity transformation is performed and the registration success is improved by 15% while the registration time improved by 23%.

### REFERENCES

[1] C. Kagadis, G. Loudos, K. Katsanos, S. Langer, and G. Nikiforidis. In vivo small animal imaging: Current status and future prospects. Medical Physics, 37(12):6421, 2010.

[2] R. Yao, R. Lecomte, and E. Crawford. Small-animal PET: what is it, and why do we need it? Journal of Nuclear Medicine Technology, 40(3):157–165, September 2012.

[3] B. Dogdas. Image registration with applications to multimodal small animal imaging, PhD Thesis. University of Southern California, 2007.

[4] B. Pichler, H. Wehrl, A. Kolb, and M.S. Judenhofer. PET / MRI : The Next Generation of Multimodality. Seminar in Nuclear Medicine, 38(3):199–208, 2008.

[5] F. Brunotte H. Haas, B. Collin, PharmD, A. Oudot, S. Bricq, A. Lalande, X. Tizon, Vrigneaud, P.M. Walker. Integrated PET/MRI in preclinical studies State of the art. tijdschrift voor nucleaire geneeskunde, 35(4):1144–1152, 2013.

[6] R. P. Woods, J. C. Mazziotta, and S. R. Cherry. MRI-PET registration with automated algorithm. Journal of Computer Assisted Tomography, 17(4):536–546, 1993.

[7] A. Collignon, F. Maes, D. Delaere, D Vandermeulen, P. Suetens, and G. Marchal. Automated multi-modality image registration based on information theory. Information Processing in Medical Imaging, pages 263–274, 1995.

[8] J. P. Pluim, J. B. Maintz, and M. Viergever. IEEE transactions on Medical Imaging, (8):809–814, August.

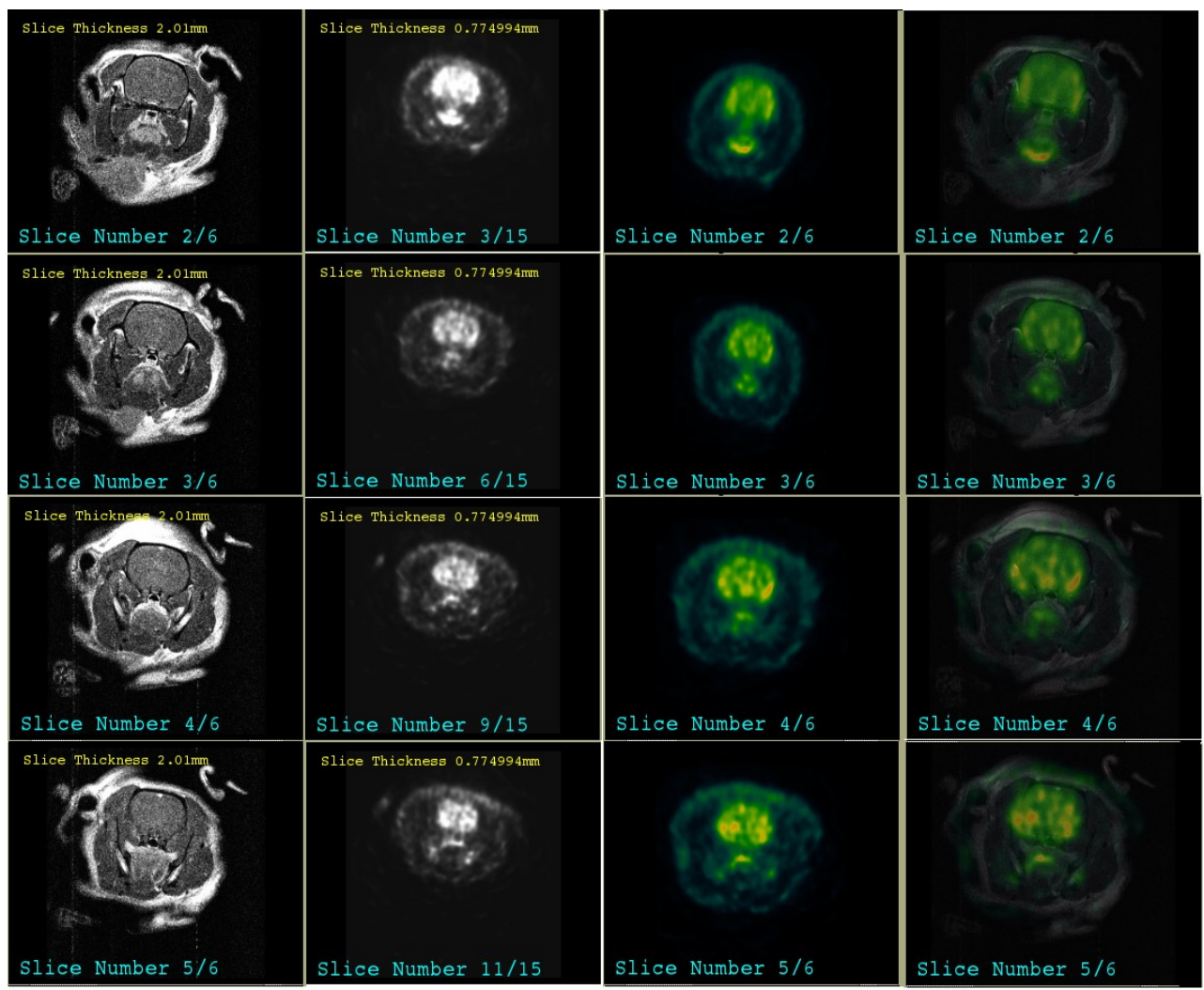

**Figure 6** Registration **successes** of the 7 datasets in %

[9] B. Ardekani, M. Braun, B. Hutton, I. Kanno, and H. Lida. A fully automatic multimodality image registration algorithm. *Journal of Computer Assisted Tomography*, 19(4):615–623, 1995.

[10] D. Mattes, D.R. Haynor, H. Vesselle, T. Lewellen, and W. Eubank. PET-CT image registration in the chest using free-form deformations. *IEEE transactions on Medical Imaging*, 22(1):120–128, January 2003.

[11] J. Vaquero and M. Desco. PET, CT, and MR image registration of the rat brain and skull. *IEEE Transactions on Nuclear Science*, 48(4):1440–1445, 2001.

[12] R. Woods, S. Cherry, and J. Mazziotta. *Journal of Computer and Tomography*, 16(4):620–635.

[13] M. Preuss, P. Werner, H. Barthel, U. Nestler, F. Wolfgang Hirsch, D. Fritzsch, M. Bernhard, and O. Sabri. A PET-MRI registration technique for PET studies of the rat brain. *Nuclear Medicine and Biology*, 27(2):121 – 125, 2000.

[14] M. Bernier, R. Lepage, L. Lecomte, L. Tremblay, L. Dor´e-Savard, and M. Descoteaux. Free-Form B-spline Deformation Model for *Groupwise Registration*. Conference Proceeding International Society of Magnetic Resonance in Medicine(ISMRM), page 3255, 2011.

[15] N.L. Baisa. MRI-PET Registration With Automated Algorithm in Preclinical Studies, VIBOT Thesis, 2013.

[16] H.J. Johnson, M. McCormick, and L. Ibanez. The ITK Software Guide. Third Edition Updated for ITK version 4.5. 2013.

[17] Saruchi. Adaptive Sigmoid Function to Enhance Low Contrast images. *International Journal of Computer Applications*, 55(4):45–49, 2012.

[18] N. M. Alpert, J. F. Bradshaw, D. Kennedy, and J. A. Correia. The Principal Axes Transformation A Method for Image Registration. *Journal of Nuclear Medicine*, 31(10):1717–1722, 1990.

[19] Z. Lu and W. Chen. Fast and Robust 3-D Image Registration Algorithm Based on Principal Component Analysis. *Bioinformatics and Biomedical Engineering*, 2007. The 1st International Conference, pages 872–875, 2007.

[20] J. Kybic and M. Unser. *IEEE transactions on image processing : a publication of the IEEE Signal Processing Society*, (11):1879–2890, January.

[21] B. Likar and F. Pernuš. A hierarchical approach to elastic registration based on mutual information. *Image and Vision Computing*, 19:33–44, 2001.

[22] J.V. Hajnal, D.L.G. Hill, and D. J. Hawkes. *Medical image registration*. CRC Press LLC, 2001.

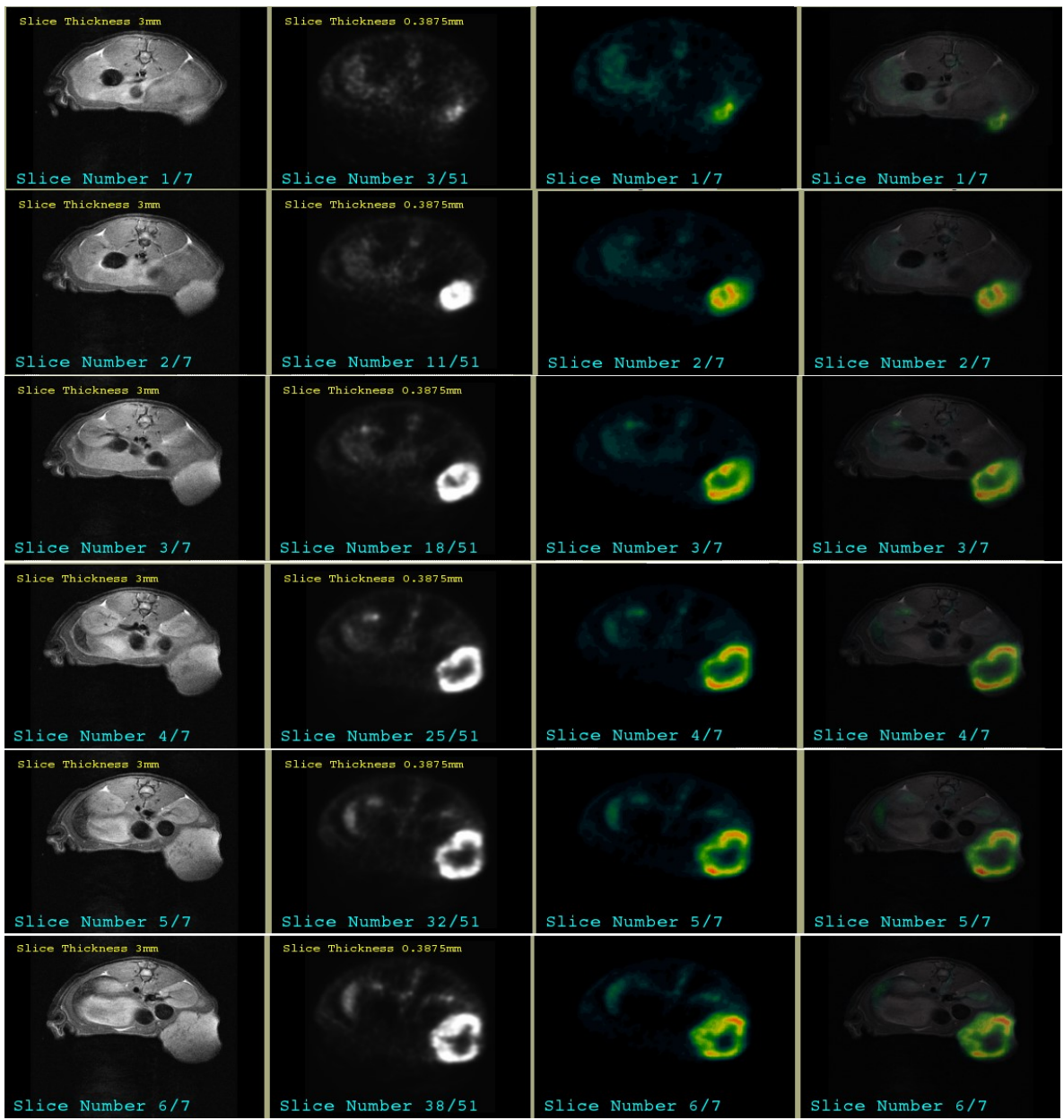

**Figure 6** Registration **successes** of the 7 datasets in %

[23] J. P. W. Pluim, J. B. A. Maintz, and M. Viergever. Mutual-informationbased registration of medical images: a survey. IEEE transactions on Medical Imaging, 22(8):986–1004, August 2003.

[24] P. Th´evenaz, T. Blu, and M. Unser. Handbook of medical imaging. chapter Image Interpolation and Resampling, pages 393–420. Academic Press, Inc., Orlando, FL, USA, 2000.

[25] C. Studholme, D. L. Hill, and D J Hawkes. Automated threedimensional registration of magnetic resonance and positron emission tomography brain images by multiresolution optimization of voxel similarity measures. Medical Physics, 24(1):25–35, January 1997.

[26] F. Maes, A. Collignon, D. Vandermeulen, G. Marchal, and P. Suetens. Multimodality image registration by maximization of mutual information. IEEE Transactions on Medical Imaging, 16(2):187–198, 1997.

[27] R. Xu, Y. Chen, S. Tang, S. Morikawa, and Y. Kurumi. Parzen-window based normalized mutual information for medical image registration. IEICE Transactions on Information and Systems, E91.D:132–144, 2010.

[28] J. Nocedal and S. J. Wright. Numerical Optimization. Springer, 1999.

[29] D. Mattes, D. R. Haynor, H. Vesselle, T. K. Lewellyn, and W. Eubank. Nonrigid multimodality image registration. SPIE Medical Imaging, pages 1609–1620, July 2001.

[30] R. Byrd, P. Lu, J. Nocedal, and C. Zhu. A limited memory algorithm for bound constrained optimization. SIAM Journal on Scientific Computing, 16(5):1190–1208, 1995.

[31] J.C. Moore. Visualizing with VTK. Linux Journal, 20:93–100, 1998.